**Quantifying the role of neurons for behavior is a mediation question.**

Ilenna Simone Jones and Konrad Paul Kording

(a1) Department of Neuroscience, University of Pennsylvania, Philadelphia, Pennsylvania, 19104, USA, ilennaj@pennmedicine.upenn.edu, http://kordinglab.com/people/ilenna_jones/index.html

(a2) Departments of Neuroscience and Bioengineering, University of Pennsylvania, Philadelphia, Pennsylvania, 19104, USA, kording@upenn.edu, http://koerding.com/

**Abstract**

Many systems neuroscientists want to understand neurons in terms of mediation; we want to understand how neurons are involved in the causal chain from stimulus to behavior. Unfortunately, most tools are inappropriate for that while our language takes mediation for granted. Here we discuss the contrast between our conceptual drive towards mediation and the difficulty of obtaining meaningful evidence.

Arguably the most popular question in systems neuroscience is about mediation: we want to know how neurons contribute to the translation from stimuli via groups of neurons to behaviors. Consequently, the field's review papers and discussion sections saliently talk about mediation as do our own papers.[1] In systems neuroscience in particular, elucidating the function or role of neurons in circuits, pathways, and networks that mediate behavior is the field's imperative. As the Brette paper points out, neurons are said to represent stimuli with which we also mean that these neurons are important to behavior.[1,2,3] Relatedly, neurons are said to encode stimuli with which we mean that they are eventually decoded and hence have a causal impact.[4–7,8,9] There are only small sections of systems neuroscience that do not primarily aim at causal descriptions. For example, Bayesian psychophysicists often do not make mechanism claims when they point out that behavior is close to optimal.[10,11] Similarly, neuroengineers trying to use brain activity to control a prosthetic device do not need to make causal assumptions.[12] But by and large, the world of ideas in systems neuroscience is a world of mediation mechanisms and algorithms, it is a world of causality.

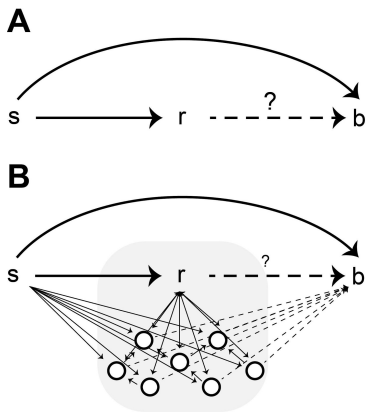

**Figure 1:** how should we think about mediation? A) if we record from all the relevant mediators then we can readily analyze mediation. B) If we only partially record activities then an analysis is very complicated.

When thinking about data, it is natural to think about the events that we are measuring. If we assume a typical recording-only experiment we have a stimulus *s*, the activity n *r* that is recorded and a behavior *b*, we can make our mediation question clear. Does r mediate the influence of *s* on *b*? We can measure the causal effect of *s* on *r*. We can also observe the relation between *s* and *b*. When only *s*, *r*, and *b* are involved, the causal inference technique called instrumental variable analysis[13] allows us to calculate the causal influence of the activity *r* onto the behavior as

$$CE(r \to b|s) = \frac{Cov(s,b)}{Cov(s,r)}$$

Where the causal effect of *r* onto *b* can be determined by the ratio of measured covariances between *s* and *b*, and between *s* and *r*. For a fixed stimulus, r will have a probability distribution (p(*r*|*s*)) and the correlation of this with *b* allows us to estimate the causal mediation effect. The necessary criterion for this reasoning is that there are no causal paths between the variables we reason about that we can not know. Importantly, if the rest of the brain does not exist, then this way of thinking about mediation analysis is perfectly good. We argue that the way we think about representations and encoding intuitively draws on this idea.

However, we typically record only a tiny proportion of all neurons.[14] Confounding then makes mediation analysis impossible. Any stimulus-behavior correlation could be due to the neurons we did record or due to other neurons that we did not. Similarly, correlations between neurons can be induced by a paired interaction or indirectly by common input from other neurons. Once our assumptions of full knowledge are violated, our estimates can be arbitrarily off, our ability to do mediation analysis is gone. We do not learn about the flow of information, or about causal chains, from the kinds of experiments popular in neuroscience.

It is possible to do experiments that get far closer to meaningful claims about mediation. There are four well established aspects that jointly make mediation more believable. (1) correlation: neurons relate to the relevant stimulus aspects and behavior (2) necessity: if the neurons are inactivated the behavior is gone. (3) sufficiency: if the neurons are activated the behavior happens. (4) exclusion: the activity is not seen in parallel streams. Such experiments are beautiful, rare[15], and generally not doable in typical mammalian settings. However, if we are after mediation effects, then these experiments should be done.[16]

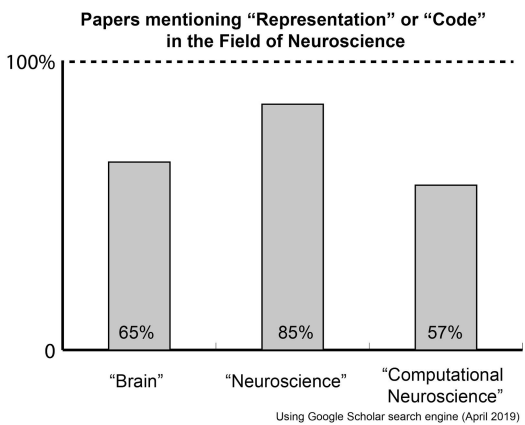

**Figure 2:** The use of Representation and Code are ubiquitous in our field.

Wordings like representation and encoding implicitly suggest that we can arrive at mediation results and, so we argue, are thus popular in neuroscience (Fig 2). Brette points out that this use of language implies that there is a causal relationship between a stimulus and an encoder, and between an encoder and an assumed decoder (as in our instrumental variable case). By consistently using words that imply mediation we are depriving the field of clarity. Language affects the way we formulate models which in turn affects the experiments we do. As such, it is not just language, but it is the core of what do as a field.

The focus of the field of mediation analysis may relate to our relative lack of real theories. Mainstream neuroscience theory subscribes to neurons influencing one another and that neurons within the same area are similar to one another. But, in a way, those kinds of theories neither do justice to the complex zoo of neural properties nor do they make the set of possible interpretations of brain data much smaller. If we had meaningful theories, we could test their predictions. Lacking theories, we then simply goes for an intuitive mediation analysis, which can not be well supported by typical experiments. Real theory, including theory that can deal with recurrent systems with circular causality, is needed to break our conceptual reliance of ideas of mediation.

Much of what we know about brains comes from the mapping of stimulus-response curves.[17] We were enabled to develop prosthetic devices[17,18] and new treatments for neurological diseases.[19] However, we should not take this impressive story of success as a sign that we do not need to clearly think about what exactly these findings mean. The focus on encoding and representation, if anything, detracts from the importance of the past findings of the field and prevents it from asking how we should think about brains. Moving forward, the field needs to invest in transcending our current theories to make real testable predictions to provide greater precision and logical power to our experiments and understanding of the brain. But tuning curves by themselves can never produce an understanding. After all, we know that theory free learning about a system is provably impossible.[20]

1. Vilares, I., Howard, J. D., Fernandes, H. L., Gottfried, J. A. & Kording, K. P. Differential representations of prior and likelihood uncertainty in the human brain. *Curr. Biol.* **22**, 1641–1648 (2012).
2. Stevenson, I. H. *et al.* Statistical assessment of the stability of neural movement representations. *J. Neurophysiol.* **106**, 764–774 (2011).
3. Körding, K. P., Kayser, C., Einhäuser, W. & König, P. How are complex cell properties adapted to the statistics of natural stimuli? *J. Neurophysiol.* **91**, 206–212 (2004).
4. Brette, R. Is coding a relevant metaphor for the brain? *Behav. Brain Sci.* 1–44 (2019).
5. Ashida, G. & Carr, C. E. Sound localization: Jeffress and beyond. *Curr. Opin. Neurobiol.* **21**, 745–751 (2011).
6. Hasselmo, M. E., Moser, E. I. & Moser, M.-B. Foreword: Special issue on grid cells. *Hippocampus* **18**, 1141–1141 (2008).
7. Jazayeri, M. & Movshon, J. A. Optimal representation of sensory information by neural populations. *Nat. Neurosci.* **9**, 690–696 (2006).
8. Klein, D. J., König, P. & Körding, K. P. Sparse Spectrotemporal Coding of Sounds. *EURASIP Journal on Advances in Signal Processing* **2003**, (2003).
9. Glaser, J. I., Perich, M. G., Ramkumar, P., Miller, L. E. & Kording, K. P. Population coding of conditional probability distributions in dorsal premotor cortex. *Nat. Commun.* **9**, 1788 (2018).
10. Körding, K. P. & Wolpert, D. M. Bayesian integration in sensorimotor learning. *Nature* **427**, 244–247

(2004).

11. Körding, K. P. & Wolpert, D. M. Bayesian decision theory in sensorimotor control. *Trends Cogn. Sci.* **10**, 319–326 (2006).

12. Wolpaw, J. R. *et al.* Brain-computer interface technology: a review of the first international meeting. *IEEE Trans. Rehabil. Eng.* **8**, 164–173 (2000).

13. Angrist, J. & Krueger, A. Instrumental Variables and the Search for Identification: From Supply and Demand to Natural Experiments. (2001). doi:10.3386/w8456

14. Stevenson, I. H. & Kording, K. P. How advances in neural recording affect data analysis. *Nat. Neurosci.* **14**, 139–142 (2011).

15. Kawashima, T., Zwart, M. F., Yang, C.-T., Mensh, B. D. & Ahrens, M. B. The Serotonergic System Tracks the Outcomes of Actions to Mediate Short-Term Motor Learning. *Cell* **167**, 933–946.e20 (2016).

16. Latimer, K. W., Yates, J. L., Meister, M. L. R., Huk, A. C. & Pillow, J. W. NEURONAL MODELING. Single-trial spike trains in parietal cortex reveal discrete steps during decision-making. *Science* **349**, 184–187 (2015).

17. Wurtz, R. H. Recounting the impact of Hubel and Wiesel. *J. Physiol.* **587**, 2817–2823 (2009).

18. Serruya, M. D., Hatsopoulos, N. G., Paninski, L., Fellows, M. R. & Donoghue, J. P. Instant neural control of a movement signal. *Nature* **416**, 141–142 (2002).

19. Perlmutter, J. S. & Mink, J. W. DEEP BRAIN STIMULATION. *Annual Review of Neuroscience* **29**, 229–257 (2006).

20. Wolpert, D. H. & Macready, W. G. No free lunch theorems for optimization. *IEEE Transactions on Evolutionary Computation* **1**, 67–82 (1997).